\documentclass{article}
\usepackage{spconf,amsmath,graphicx}

\usepackage{amssymb,amsfonts}
\usepackage[capitalise, noabbrev]{cleveref}
\usepackage{url}
\usepackage{siunitx}
\usepackage{booktabs}

\title{Listening Between the Lines:\\ Synthetic Speech Detection Disregarding Verbal Content}

\name{Davide Salvi, Temesgen Semu Balcha, Paolo Bestagini, Stefano Tubaro
\thanks{This material is based on research sponsored by the Defense Advanced Research Projects Agency (DARPA) and the Air Force Research Laboratory (AFRL) under agreement number FA8750-20-2-1004. The U.S. Government is authorized to reproduce and distribute reprints for Governmental purposes notwithstanding any copyright notation thereon. The views and conclusions contained herein are those of the authors and should not be interpreted as necessarily representing the official policies or endorsements, either expressed or implied, of DARPA and AFRL or the U.S. Government.
This work was supported by the FOSTERER project, funded by the Italian Ministry of Education, University, and Research within the PRIN 2022 program. This work was partially supported by the European Union under the Italian National Recovery and Resilience Plan (NRRP) of NextGenerationEU, partnership on ``Telecommunications of the Future'' (PE00000001 - program ``RESTART'').}}
\address{Dipartimento di Elettronica, Informazione e Bioingegneria, Politecnico di Milano - Milan, Italy\\
{\tt \small\{davide.salvi, paolo.bestagini, stefano.tubaro\}@polimi.it}}

\usepackage{glossaries}
\newacronym{xai}{XAI}{Explainable AI}
\newacronym{shap}{SHAP}{SHapley Additive exPlanations}
\newacronym{gradcam}{Grad-CAM}{Gradient-weighted Class Activation Mapping}
\newacronym{lime}{LIME}{Local Interpretable Model-agnostic Explanations}
\newacronym{gru}{GRU}{Gated Recurrent Unit}
\newacronym{roc}{ROC}{Receiver Operating Characteristic}
\newacronym{auc}{AUC}{Area Under the Curve}

\begin{document}
\ninept

\maketitle

\begin{abstract}
Recent advancements in synthetic speech generation have led to the creation of forged audio data that are almost indistinguishable from real speech. This phenomenon poses a new challenge for the multimedia forensics community, as the misuse of synthetic media can potentially cause adverse consequences.
Several methods have been proposed in the literature to mitigate potential risks and detect synthetic speech, mainly focusing on the analysis of the speech itself.
However, recent studies have revealed that the most crucial frequency bands for detection lie in the highest ranges (above 6000 Hz), which do not include any speech content.
In this work, we extensively explore this aspect and investigate whether synthetic speech detection can be performed by focusing only on the background component of the signal while disregarding its verbal content.
Our findings indicate that the speech component is not the predominant factor in performing synthetic speech detection. 
These insights provide valuable guidance for the development of new synthetic speech detectors and their interpretability, together with some considerations on the existing work in the audio forensics field.
\end{abstract}


\section{Introduction}
\label{sec:introduction}

In the last few years, the scientific community has made significant advancements in the synthetic speech generation field.
It is now possible to generate highly realistic speech tracks that mimic a target speaker's voice using commonplace devices and open-access tools that operate with minimal computational resources.
This progress proves advantageous in contexts like human-machine interaction and accessibility for individuals with speech disorders~\cite{yamagishi2012speech}. 
Nonetheless, the ability to generate synthetic speech also introduces ethical concerns, particularly regarding its potential misuse for malicious purposes~\cite{NYT_audio}.
This is the case of audio deepfakes, synthetic speech signals produced through AI-driven technologies that can clone a target speaker's voice, making them utter statements they never made.
The potential risks associated with this phenomenon have prompted the multimedia forensics community to address this concern actively, working on developing systems capable of verifying the authenticity of audio files~\cite{cuccovillo2022open}.
Several synthetic speech detection techniques have been introduced employing a diverse range of approaches, including end-to-end methods~\cite{tak2021end, ma2023end}, and systems based on the analysis of acoustic~\cite{hamza2022deepfake}, and semantic features~\cite{conti2022deepfake, attorresi2022prosody}.
The proposed detectors leverage cutting-edge technologies in the AI field and show remarkable performance, especially in controlled scenarios. 
However, due to their data-driven nature, they are often used as \textit{black-boxes} with limited interpretability. This constraint restricts their application in real-world scenarios where a comprehensive understanding of what is driving the detection process is essential.

To overcome this limitation, the scientific community has increased its attention towards \gls{xai}, aiming to understand the critical elements in an audio track that influence the predictions of a synthetic speech detector.
For instance, the authors of~\cite{ge2022explaining} and~\cite{ge2022explainable} employ the \gls{shap} method to analyze the artifacts produced by synthetic speech generators, while those of~\cite{halpern123residual} and~\cite{chettri2018analysing} utilize GradCAM and LIME algorithms to gain insight into the decision-making process of synthetic speech detectors, respectively. 
Finally, other studies~\cite{takexplainability},
have focused on explaining the results obtained using specific audio features, focusing on different frequency bands and speech formants.

In our latest study~\cite{salvi2023towards}, we investigated which are the most critical frequency bands to perform synthetic speech detection.
We found that the most significant one is located in the highest frequency range (above \SI{6000}{\hertz}), characterized by the absence of any verbal content and comprising only background noise.
This insight is crucial, as many detectors are specifically designed to analyze speech nuances while disregarding the noise component.
Realizing that the meaningful content for this task lies in the background allows us to develop detectors that are informed of this fact, resulting in enhanced performance and reliability.
These findings are also motivated by the fact that the analysis of high-frequency components is a well-established technique in multimedia forensics, with this approach already being used for image processing and image splicing detection~\cite{cozzolino2015splicebuster}.
However, to our knowledge, this methodology has never been employed in the synthetic speech detection field, where its potential value could be considerable.

In this paper we aim to deepen this investigation. 
Starting from an input speech track, we separate its verbal and background noise components and use them individually for synthetic speech detection.
Our findings indicate that analyzing the background noise alone leads to better classification results across diverse scenarios.
Furthermore, we show how the results obtained are not significant only for one specific noise extractor but can also be extended to other methods, providing valuable insights into the generalization capabilities of the proposed method.
 
\section{Synthetic Speech Detection}
\label{sec:method}

In this paper we consider the problem of synthetic speech detection and investigate whether the signal components that are most relevant for this task are predominantly present in the speech or the background noise.
In our previous work~\cite{salvi2023towards}, we showed that the most crucial frequency band for detection is the highest one (above \SI{6000}{\hertz}), which does not include any verbal content.
Starting from these results, we aim to study whether it is possible to discriminate between real and fake speech signals solely by looking at the background noise component and disregarding any verbal information.
Previous studies in the field have explored synthetic speech detection by analyzing only the silent portions of the speech signal, leading to considerations regarding what we really need to analyze in an audio track~\cite{muller2021speech, mari2022sound}.

The rationale behind this work is that synthetic speech generators have been trained to replicate the voice of a target speaker, producing high-quality results on this task.
However, due to their emphasis on this specific element, they may overlook the synthesis of other aspects of the signal, such as background noise and high-frequency components, which may include some artifacts.
Therefore, we can leverage this aspect and perform a more accurate synthetic speech detection by focusing only on these components.

\subsection{Problem Formulation}

The synthetic speech detection problem is formally defined as follows.
Given a discrete-time input speech signal $\mathbf{x}$ sampled with sampling frequency $f_\text{s}$, the goal is to build a detector $\mathcal{D}$ able to predict the class $y \in \{Real, Fake \}$ associated to $\mathbf{x}$.
Here, \textit{Real} indicates authentic speech, while \textit{Fake} means that the signal has been generated through any speech synthesis technique.
We represent the speech signal $\mathbf{x}$ as the sum of two different components, such as 
\begin{equation}
    \mathbf{x} = \mathbf{s} + \mathbf{n} ,
\end{equation}
where $\mathbf{s}$ and $\mathbf{n}$ are the speech (or verbal) and noise components present in the audio track, respectively.

Our proposed approach extends beyond the standard synthetic speech detection on the entire signal $\mathbf{x}$.
Instead, we train three distinct synthetic speech detectors, denoted as $\mathcal{D}_\mathbf{x}$, $\mathcal{D}_\mathbf{s}$, and $\mathcal{D}_\mathbf{n}$, which take as input the signals $\mathbf{x}$, $\mathbf{s}$, and $\mathbf{n}$, respectively, and output 
an estimate of the class $y$ of the signal based on the input they receive.
We aim to assess and compare the performance of the three detectors and determine which one leads to the most accurate predictions.
Our findings will contribute to a deeper understanding of synthetic speech generation and detection tasks, helping the development of more accurate detectors in the future.

\subsection{Proposed Pipeline}
\label{sec:pipeline}
\begin{figure}
  \centering 
  \includegraphics[width=.7\columnwidth]{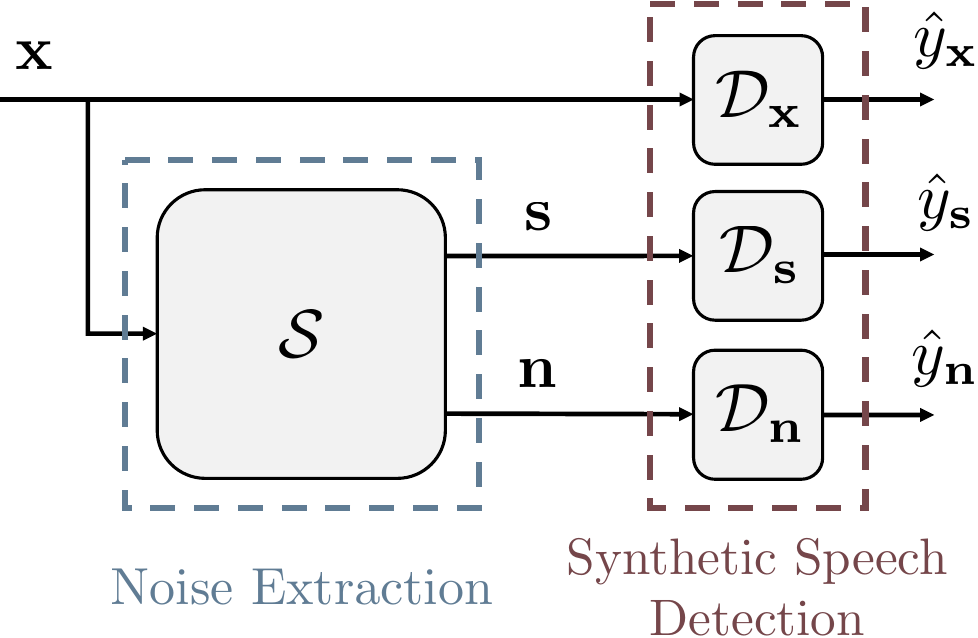}
  \caption{Pipeline of the proposed method.}
  \label{fig:pipeline}
  \vspace{-1em}
\end{figure}

The proposed pipeline consists of two distinct steps.
The first one is the noise extraction stage and involves the speech separator module $\mathcal{S}$, which extracts the estimates of the two components $\mathbf{s}$ and $\mathbf{n}$ from the input signal $\mathbf{x}$, such that 
\begin{equation}
    \mathbf{s}, \mathbf{n} = \mathcal{S}(\mathbf{x}).
    \label{eq:speech_sep}
\end{equation}
With a slight abuse of notation, since we do not have access to the original verbal and noise components of the signal $\mathbf{x}$, we refer to their estimates as $\mathbf{s}$ and $\mathbf{n}$.
The second step is the synthetic speech detection stage and involves the detectors $\mathcal{D}_i$, which takes as input the signal $i$ and outputs the class estimate $\hat{y}_i$, such that
\begin{equation}
    \hat{y}_i = \mathcal{D}_i(i) , 
\end{equation}
with $i \in \{\mathbf{x}, \mathbf{s}, \mathbf{n} \}$.
We highlight that each detector $\mathcal{D}_i$ is trained and tested only on one type of signal (i.e., original $\mathbf{x}$, speech-only $\mathbf{s}$, noise-only $\mathbf{n}$).
\cref{fig:pipeline} shows a block diagram of the proposed pipeline.

\subsubsection{Noise Extraction Stage}
\label{subsec:noise_extr}

The noise extraction stage of the pipeline receives as input the original signal $\mathbf{x}$ and outputs its speech and noise components $\mathbf{s}$ and $\mathbf{n}$.
We implement this stage using two different approaches to evaluate the generalization capabilities of the proposed method and demonstrate that it is not reliant on a specific noise extractor system.

The first noise extractor we consider is the speech enhancement model introduced in~\cite{defossez2020real}.
This follows an encoder-decoder architecture with skip connections that return a speech-enhanced version of the input signal.
We utilize a version of the model trained on the Deep Noise Suppression (DNS) Challenge dataset, presented at Interspeech 2020~\cite{reddy2020interspeech}. 
We refer to this model as \textit{Demucs} $\mathcal{S}_\text{DMCS}$.
%
The second approach we consider is \textit{SepFormer}, a speech separation model presented in~\cite{subakan2021attention}. This system is trained on the WHAMR! dataset~\cite{Maciejewski2020WHAMR} to separate speech from background noise. 
Both the models $\mathcal{S}_\text{DMCS}$ and $\mathcal{S}_\text{SF}$ take as input the original signal $\mathbf{x}$ and directly output the two components $\mathbf{s}$ and $\mathbf{n}$, as indicated in \cref{eq:speech_sep}.
  
\subsubsection{Synthetic Speech Detection Stage}
\label{subsec:detection}

The synthetic speech detector $\mathcal{D}_i$ we consider in this study is RawNet2~\cite{tak2021end}.
This is an end-to-end neural network for audio deepfake detection that directly operates on raw waveform inputs.
Originally presented in the ASVspoof 2019 challenge~\cite{todisco2019asvspoof} and included as a baseline in the ASVspoof 2021 challenge~\cite{yamagishi2021asvspoof}, RawNet2 features Sinc filters from SincNet~\cite{ravanelli2018speaker} followed by two Residual Blocks with skip connections on a GRU layer, aiming to extract frame-level representations of the input signal.
For our study, we use the exact architecture proposed in the original paper, with the only modification being the input size of the network from \SI{4}{\second} to \SI{2}{\second}. This adjustment makes the model more suitable to address the latest challenges in synthetic speech detection, where there is a need to have more punctual predictions of the authenticity of the signal over time to detect counterfeit media that interleave real and fake speech segments.

We consider the same model for all three detectors ($\mathcal{D}_\mathbf{x}$, $\mathcal{D}_\mathbf{s}$ and $\mathcal{D}_\mathbf{n}$), with the only distinctions being the input signal and the corresponding training strategy.
This choice ensures that the differences in performance of the models depend solely on the content of the signals and allows us to understand which of them contains more relevant information for the synthetic speech detection task.

\begin{figure*}[t]
    \centering
    \includegraphics[width=0.3\textwidth]{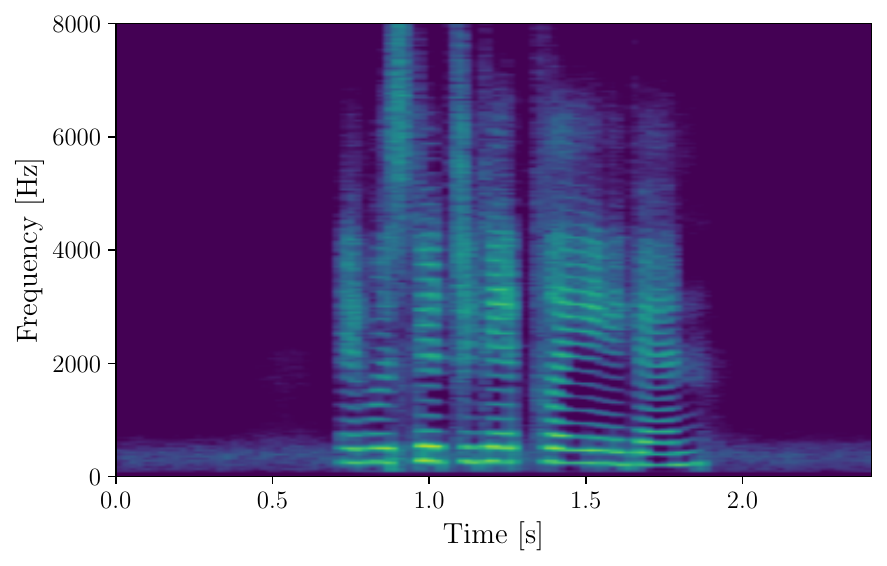} \quad
    \includegraphics[width=0.3\textwidth]{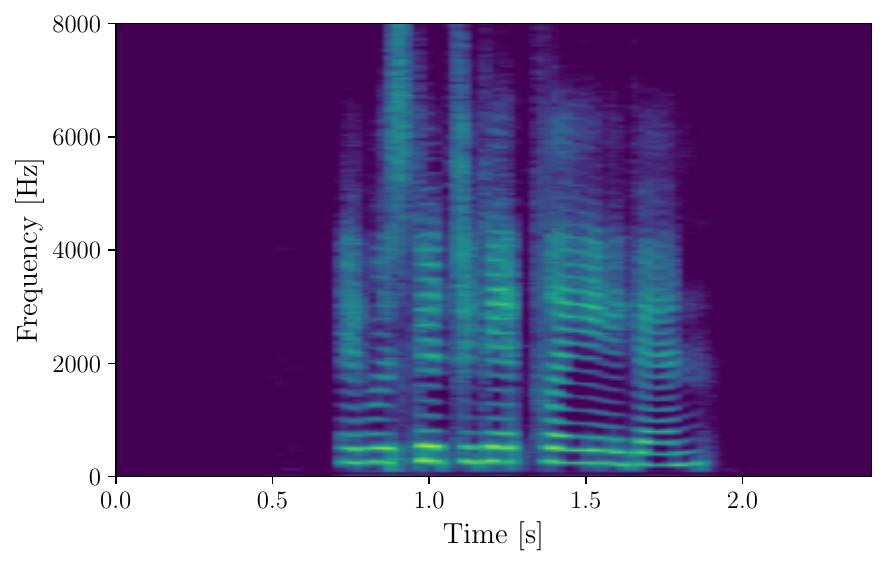} \quad
    \includegraphics[width=0.3\textwidth]{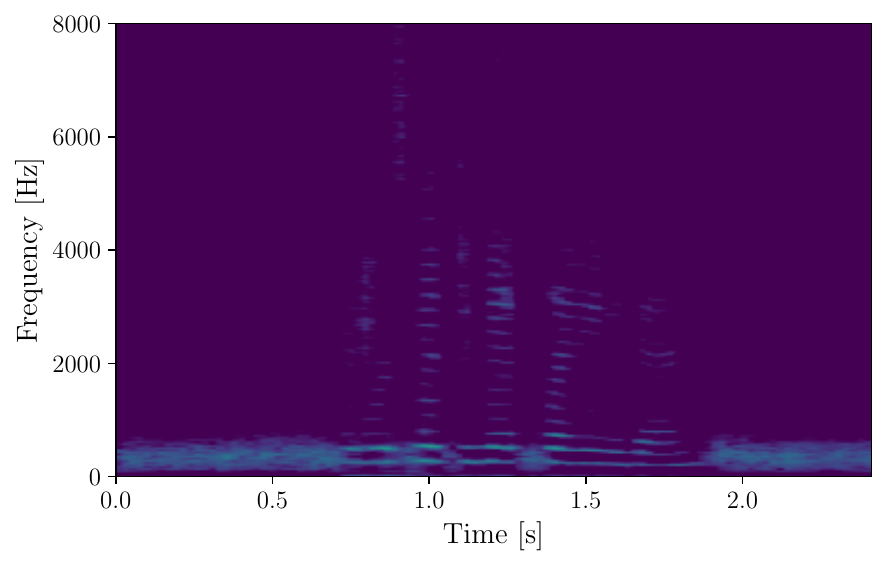}
    \caption{Spectrograms of the signals $\textbf{x}$ (left), $\textbf{s}$ (center) and $\textbf{n}$ (right) of an example track. The two tracks  $\textbf{s}$ and  $\textbf{n}$ have been computed using the $\mathcal{S}_\text{DMCS}$ model.}
    \label{fig:spectrograms}
    \vspace{-1.0em}
\end{figure*}

\section{Experimental Setup}
\label{sec:setup}



During all the experiments, we use RawNet2 as synthetic speech detector, following the implementation proposed in~\cite{tak2021end}.
As anticipated in \cref{subsec:detection}, we reduce the input length of the network from \SI{4}{\second} to \SI{2}{\second}, translating to a change from \num{64000} to \num{32000} samples, considering a sampling frequency equal to $f_\text{s}$ = \SI{16}{\kilo\hertz}.
We follow the same training strategy for all $\mathcal{D}_\textbf{x}$, $\mathcal{D}_\textbf{s}$ and $\mathcal{D}_\textbf{n}$.
The networks have been trained for \num{150} epochs with an early stopping of \num{15} epochs, considering a batch size of \num{128} samples and a learning rate of $10^{-4}$. We assumed Cross Entropy as loss function with a label smoothing equal to \num{0.2}.
During training, we balanced each batch with equal samples for the \textit{Real} and \textit{Fake} classes.


The dataset used to train and test the models is ASVspoof 2019~\cite{todisco2019asvspoof}, a speech audio set created to develop antispoofing techniques for automatic speaker verification. We consider the Logical Access (LA) partition of the dataset, which contains both real and fake data generated with several synthesis techniques. The speech generation algorithm employed in the training and validation partitions of this dataset differs from the one utilized in the test partition, allowing the testing of the synthetic speech detectors in an open-set scenario.

To assess the generalization capabilities proposed approach, we test the models also on other datasets released in the literature.
These are ASVspoof 2021~\cite{yamagishi2021asvspoof}, AISEC ``In-the-Wild''~\cite{muller2022does}, and FakeOrReal~\cite{reimao2019dataset}.
The reason why we do so is to verify whether our findings are restricted to a single dataset or can be generalized to more \textit{in the wild} conditions. This is a crucial aspect in multimedia forensics, where we want our detectors to be as performant as possible even in conditions that are different from the ones seen during training.

We also want to test the robustness of the proposed detectors against anti-forensics attacks, i.e., operations applied to the signals to degrade their quality and compromise the accuracy of deepfake detection systems.
In particular, we consider an MP3 compression attack, a processing operation that is commonly applied to speech tracks on the web or social platforms.
We also want to test this particular attack because it directly affects the high-frequency components of audio signals, potentially impacting the performance of the proposed $\mathcal{D}_\textbf{n}$ detector.
We simulate this attack using the Python \textit{audiomentations} library, considering three different compression values (\textit{bitrate} [\si{\kilo \bit / \second}] = [192, 128, 64]), applied prior to the noise extraction stage.

During all the experiments, we train and test the detector $\mathcal{D}_\mathbf{x}$ on the original audio files released in the datasets, while the detectors $\mathcal{D}_\mathbf{s}$ and $\mathcal{D}_\mathbf{n}$ are trained on the signals $\mathbf{s}$ and $\mathbf{n}$, extracted from $\mathbf{x}$ using one between the $\mathcal{S}_\text{DMCS}$ and $\mathcal{S}_\text{SF}$ models presented in \cref{subsec:noise_extr}.
$\mathcal{D}_\textbf{x}$ also serves as a baseline for our experiments, as it is a state-of-the-art detector trained and tested under the same conditions for which it was originally designed.

\section{Results}
\label{sec:results}

In this section, we evaluate the effectiveness of the proposed approach.
We aim to determine whether performing synthetic speech detection considering only one of the two signal components, either $\mathbf{s}$ or $\mathbf{n}$, leads to better results compared to using the original signal $\mathbf{x}$.
Figure~\ref{fig:spectrograms} shows the spectrograms of the three signals, computed using $\mathcal{S}_\text{DMCS}$ as noise extractor.
We assess the performance of the models across various conditions, investigating their generalization capabilities on unseen data and robustness to MP3 compression.

\begin{figure}[!b]
    \centering
    \vspace{-0.5em}
    \includegraphics[width=.6\columnwidth]{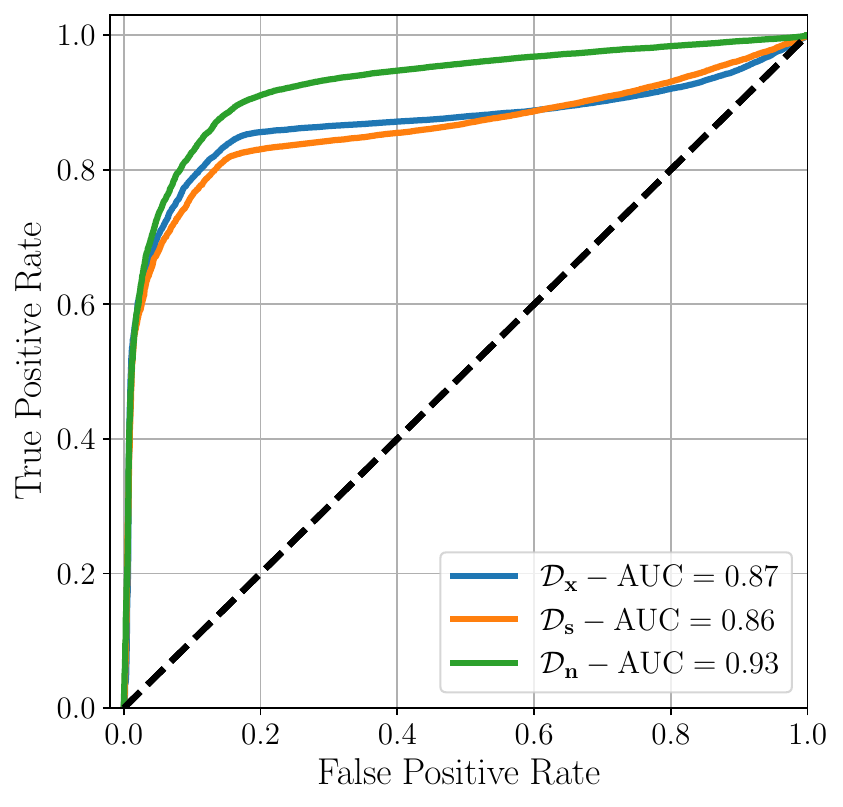}
    \vspace{-1.0em}
    \caption{ROC curve showing the synthetic speech detection performances of $\mathcal{D}_\mathbf{x}$, $\mathcal{D}_\mathbf{s}$ and $\mathcal{D}_\mathbf{n}$ tested on the ASVspoof 2019 dataset, considering $\mathcal{S}_\text{DMCS}$ as noise extractor.}
    \label{fig:DNS_ROC}
\end{figure}

\vspace{0.5em}
\noindent
\textbf{Detection Results.}
In our initial experiment, we evaluate the performance of the proposed pipeline by training and testing the detectors on the ASVspoof 2019 dataset.
\cref{fig:DNS_ROC} show the results of this experiment through \gls{roc} curves, considering $\mathcal{S}_\text{DMCS}$ as noise extractor.
$\mathcal{D}_\mathbf{n}$ consistently outperforms the other two detectors, showcasing an \gls{auc} improvement of \num{0.06} compared to the baseline $\mathcal{D}_\mathbf{x}$.

As outlined in \cref{sec:method}, our analysis extends beyond a single noise extractor. We explore two different approaches to illustrate that the obtained results are not confined to a specific model but rather can be generalized.
Table~\ref{tab:metrics_separators} presents the \gls{auc} and balanced accuracy values of the detectors when considering both $\mathcal{S}_\text{DMCS}$ and $\mathcal{S}_\text{SF}$ as noise extractors.
In both cases, the best results are achieved by employing the detector $\mathcal{D}_\mathbf{n}$ that works only on the background noise signal $\textbf{n}$.
This supports our initial hypothesis, emphasizing that the primary information for synthetic speech detection predominantly resides in the background noise rather than the speech itself.

Given that the pipeline featuring $\mathcal{S}_\text{DMCS}$ achieves superior results, we exclusively present the results obtained with this model in the following experiments.

\begin{table}
\centering
\caption{AUC and Balanced Accuracy values of $\mathcal{D}_\mathbf{x}$, $\mathcal{D}_\mathbf{s}$ and $\mathcal{D}_\mathbf{n}$ tested on the ASVspoof 2019 dataset, considering both $\mathcal{S}_\text{DMCS}$ and $\mathcal{S}_\text{SF}$ as noise extractors.}
\vspace{0.75em}
\label{tab:metrics_separators}
\begin{tabular}{@{\hspace{.25cm}} c @{\hspace{.85cm}} ccccc}
    \toprule
   & \multicolumn{2}{c}{$\mathcal{S}_\text{DMCS}$} & & \multicolumn{2}{c}{$\mathcal{S}_\text{SF}$} \\
   & AUC       & B. ACC.     & & AUC       & B. ACC.     \\ \midrule
$\mathcal{D}_\mathbf{x}$ & \num{0.87} & \num{0.82} & & \num{0.87} &  \textbf{\num{0.82}} \\
$\mathcal{D}_\mathbf{s}$ & \num{0.86} & \num{0.81} &&  \num{0.89} &  \num{0.79} \\
$\mathcal{D}_\mathbf{n}$ & \textbf{\num{0.93}} & \textbf{\num{0.83}} & &  \textbf{\num{0.90}} &  \textbf{\num{0.82}} \\ \bottomrule
\end{tabular}
\vspace{-1em}
\end{table}


\vspace{0.5em}
\noindent
\textbf{Generalization Capabilities.}
Given the results of the proposed approach when tested on ASVspoof 2019, we want to extend the analysis by assessing its performance on datasets other than the one seen in training.
This experiment aims to investigate whether our findings are dataset-specific or can be generalized.
\cref{tab:external_dataset} presents the obtained results employing balanced accuracy as a metric. To compute these scores, we threshold the softmax output of the network at \num{0.5}.
In this experiment, $\mathcal{D}_\mathbf{x}$ serves as a baseline, representing the state-of-the-art without influence from our proposed method.
In contrast, the performances of $\mathcal{D}_\mathbf{s}$ and $\mathcal{D}_\mathbf{n}$ models represent the outcomes of our pipeline.

Across all datasets examined, the $\mathcal{D}_\mathbf{n}$ model consistently outperforms the other two, exhibiting accuracy improvements ranging from \num{2}\% in the case of ASVspoof 2021 to \num{35}\% when considering the FakeOrReal dataset, compared to the baseline $\mathcal{D}_\mathbf{x}$.
While acknowledging that the performance of $\mathcal{D}_\mathbf{x}$ may not be exceptionally high, it represents the state-of-the-art benchmark and, for this reason, we assume it as suitable for comparison.
Given that our proposed pipeline addresses only track preprocessing without impacting the detector itself, we expect that our results can also be generalized to more performing detectors.

Finally, the performance of $\mathcal{D}_\mathbf{x}$ and $\mathcal{D}_\mathbf{s}$ are comparable, suggesting that the $\mathcal{D}_\mathbf{x}$ detector relies its predictions primarily on the verbal content of the signal.

\begin{figure}
    \centering
    \vspace{1em}
    \includegraphics[width=.8\columnwidth]{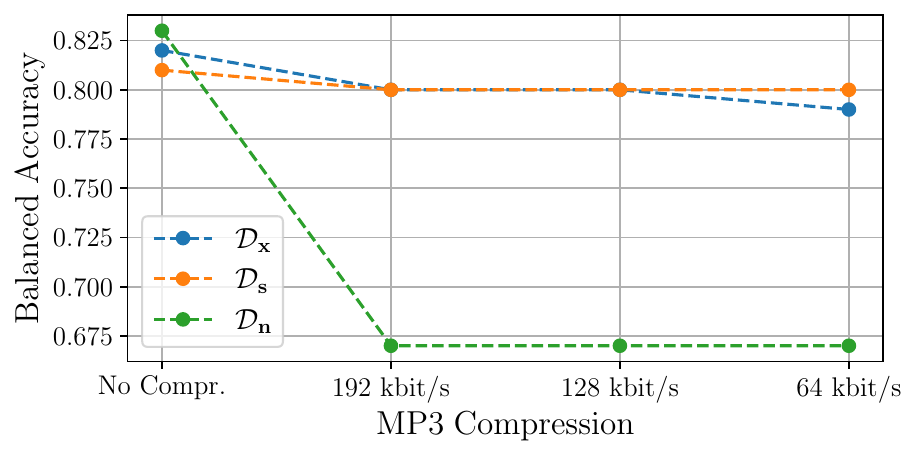}
    \caption{Balanced accuracy values scored by $\mathcal{D}_\mathbf{x}$, $\mathcal{D}_\mathbf{s}$ and $\mathcal{D}_\mathbf{n}$ tested on the ASVspoof 2019 dataset under varying MP3 compression bitrates.}
    \label{fig:mp3_compression}
\end{figure}

\vspace{0.5em}
\noindent
\textbf{Compression Robustness.}
As a final experiment, we test the robustness of the proposed pipeline against anti-forensics attacks.
Synthetic speech signals typically preserve traces and artifacts from their generators, which can be exploited for their discrimination. However, post-processing operations may distort or eliminate these traces, making the detection process more challenging.
When such post-processing operations are employed intentionally to inhibit forensic detectors, we refer to them as anti-forensics attacks.
We consider MP3 compression as an attack and examine its impact on detector performance, testing the model on ASVspoof 2019 at various compression bitrates.

\cref{fig:mp3_compression} shows the balanced accuracy values achieved in the considered cases.
$\mathcal{D}_\mathbf{s}$ proves to be the most robust method of the three, exhibiting no tangible impact from the post-processing operation. $\mathcal{D}_\mathbf{x}$ achieves acceptable results, with an accuracy drop of only \num{3}\%.
In contrast, $\mathcal{D}_\mathbf{n}$ exhibits a notable performance decrease, with a \num{15}\% drop.
This result is somehow expected, as MP3 compression primarily affects the high frequencies and the non-speech content, upon which $\mathcal{D}_\mathbf{n}$ bases its predictions. 
These results underscore the importance of considering such factors in developing or training synthetic speech detectors to prevent them from any bias.

\begin{table}
\centering
\caption{Balanced accuracy values scored by $\mathcal{D}_\mathbf{x}$, $\mathcal{D}_\mathbf{s}$ and $\mathcal{D}_\mathbf{n}$ tested on unseen datasets during training.}
\vspace{0.75em}
\label{tab:external_dataset}
\begin{tabular}{l @{\hspace{.75cm}} ccc}
\toprule
  & $\mathcal{D}_\mathbf{x}$ & $\mathcal{D}_\mathbf{s}$ & $\mathcal{D}_\mathbf{n}$ \\ \midrule
ASVspoof 2019 \cite{todisco2019asvspoof} & \num{0.82} & \num{0.81} & \textbf{\num{0.83}} \\
ASVspoof 2021 \cite{yamagishi2021asvspoof} & \num{0.72} & \num{0.72} & \textbf{\num{0.73}} \\
FakeOrReal \cite{reimao2019dataset} & \num{0.53} & \num{0.54} & \textbf{\num{0.71}} \\
AISEC "In-the-Wild" \cite{muller2022does} & \num{0.53} & \num{0.53} & \textbf{\num{0.69}} \\
ADD 2022 \textit{train} \cite{yi2022add} & \num{0.48} & \num{0.44} & \textbf{\num{0.54}} \\
\bottomrule
\end{tabular}
\end{table}

\vspace{0.5em}
\noindent
\textbf{Discussion.}
The obtained results provide a basis for comprehensive discussions on multiple fronts. 
First, the superior performance of $\mathcal{D}_\mathbf{n}$ compared to $\mathcal{D}_\mathbf{s}$ prompts reflection about \gls{xai} in synthetic speech detection and the specific aspects of the audio signals that the detectors analyze to that drive their predictions. While common intuition might suggest that detectors primarily focus on verbal content, our findings indicate this is not the case.
Secondly, the consistency of this trend across various datasets, combined with the limited generalization capabilities of the detectors, makes us reflect on the quality of the data and the models released in the state-of-the-art for synthetic speech detection.
This suggests the presence of potential dataset-specific characteristics in the data or that the detectors may be prone to overfitting on specific datasets, limiting their generalization capabilities.
Finally, the results of $\mathcal{D}_\mathbf{n}$ against anti-forensic attacks raise concerns about the suitability of detectors that exclusively analyze specific signal components for real-world scenarios.
This leads to considerations about the feasibility of implementing informed detectors capable of analyzing the distinct components of the signal differently, extracting the most pertinent information from each of them.
In a broader context, we advocate for the development of synthetic speech detectors under realistic conditions, enhancing their generalization capabilities across various datasets and their robustness against post-processed data.
Prioritizing real-world applicability is crucial for advancing the field and ensuring the practical effectiveness of synthetic speech detection systems.

\section{Conclusion}
\label{sec:conclusion}

In this paper, we addressed the problem of synthetic speech detection and investigated the possibility of performing this task by exclusively examining the background component of the signal while disregarding its verbal content. 
Utilizing two distinct noise extraction methods, we validated our initial hypothesis and assessed its generalization capabilities and robustness to MP3 compression.
The obtained results led to interesting considerations regarding  \gls{xai} in synthetic speech detection, the quality of state-of-the-art data, and the approaches used in synthetic speech detectors.
Future research will focus on enhancing the pipeline's robustness against anti-forensic attacks and exploring more effective implementation approaches.

\newpage

\bibliographystyle{IEEEtran}
\bibliography{bstcontrol, refs}

\begin{thebibliography}{10}
\providecommand{\url}[1]{#1}
\csname url@samestyle\endcsname
\providecommand{\newblock}{\relax}
\providecommand{\bibinfo}[2]{#2}
\providecommand{\BIBentrySTDinterwordspacing}{\spaceskip=0pt\relax}
\providecommand{\BIBentryALTinterwordstretchfactor}{4}
\providecommand{\BIBentryALTinterwordspacing}{\spaceskip=\fontdimen2\font plus
\BIBentryALTinterwordstretchfactor\fontdimen3\font minus \fontdimen4\font\relax}
\providecommand{\BIBforeignlanguage}[2]{{%
\expandafter\ifx\csname l@#1\endcsname\relax
\typeout{** WARNING: IEEEtran.bst: No hyphenation pattern has been}%
\typeout{** loaded for the language `#1'. Using the pattern for}%
\typeout{** the default language instead.}%
\else
\language=\csname l@#1\endcsname
\fi
#2}}
\providecommand{\BIBdecl}{\relax}
\BIBdecl

\bibitem{yamagishi2012speech}
J.~Yamagishi, C.~Veaux, S.~King, and S.~Renals, ``Speech synthesis technologies for individuals with vocal disabilities: Voice banking and reconstruction,'' \emph{Acoustical Science and Technology}, vol.~33, no.~1, pp. 1--5, 2012.

\bibitem{NYT_audio}
N.~Y. Times, ``{‘A.I. Obama’ and Fake Newscaster: How A.I. Audio is Swarming TikTok},'' \url{https://www.nytimes.com/2023/10/12/technology/tiktok-ai-generated-voices-disinformation.html}.

\bibitem{cuccovillo2022open}
L.~Cuccovillo, C.~Papastergiopoulos, A.~Vafeiadis, A.~Yaroshchuk, P.~Aichroth, K.~Votis, and D.~Tzovaras, ``Open challenges in synthetic speech detection,'' in \emph{IEEE International Workshop on Information Forensics and Security (WIFS)}, 2022.

\bibitem{tak2021end}
H.~Tak, J.~Patino, M.~Todisco, A.~Nautsch, N.~Evans, and A.~Larcher, ``{End-to-end anti-spoofing with RawNet2},'' in \emph{IEEE International Conference on Acoustics, Speech and Signal Processing (ICASSP)}, 2021.

\bibitem{ma2023end}
K.~Ma, Y.~Feng, B.~Chen, and G.~Zhao, ``End-to-end dual-branch network towards synthetic speech detection,'' \emph{IEEE Signal Processing Letters}, vol.~30, pp. 359--363, 2023.

\bibitem{hamza2022deepfake}
A.~Hamza, A.~R.~R. Javed, F.~Iqbal, N.~Kryvinska, A.~S. Almadhor, Z.~Jalil, and R.~Borghol, ``Deepfake audio detection via mfcc features using machine learning,'' \emph{IEEE Access}, vol.~10, pp. 134\,018--134\,028, 2022.

\bibitem{conti2022deepfake}
E.~Conti, D.~Salvi, C.~Borrelli, B.~Hosler, P.~Bestagini, F.~Antonacci, A.~Sarti, M.~C. Stamm, and S.~Tubaro, ``{Deepfake Speech Detection Through Emotion Recognition: a Semantic Approach},'' in \emph{IEEE International Conference on Acoustics, Speech and Signal Processing (ICASSP)}, 2022.

\bibitem{attorresi2022prosody}
L.~Attorresi, D.~Salvi, C.~Borrelli, P.~Bestagini, and S.~Tubaro, ``{Combining Automatic Speaker Verification and Prosody Analysis for Synthetic Speech Detection},'' in \emph{International Conference on Pattern Recognition (ICPR)}, 2022.

\bibitem{ge2022explaining}
W.~Ge, J.~Patino, M.~Todisco, and N.~Evans, ``{Explaining deep learning models for spoofing and deepfake detection with SHapley Additive exPlanations},'' in \emph{IEEE International Conference on Acoustics, Speech and Signal Processing (ICASSP)}.\hskip 1em plus 0.5em minus 0.4em\relax IEEE, 2022.

\bibitem{ge2022explainable}
W.~Ge, M.~Todisco, and N.~Evans, ``{Explainable deepfake and spoofing detection: an attack analysis using SHapley Additive exPlanations},'' \emph{arXiv preprint arXiv:2202.13693}, 2022.

\bibitem{halpern123residual}
B.~M. Halpern123, F.~Kelly, R.~van Son12, and A.~Alexander, ``Residual networks for resisting noise: analysis of an embeddings-based spoofing countermeasure,'' in \emph{Speaker and Language Recognition Workshop (Odyssey)}, 2020.

\bibitem{chettri2018analysing}
B.~Chettri, S.~Mishra, B.~L. Sturm, and E.~Benetos, ``{Analysing the predictions of a CNN-based replay spoofing detection system},'' in \emph{IEEE Spoken Language Technology Workshop}, 2018.

\bibitem{takexplainability}
H.~Tak, J.~Patino, A.~Nautsch, N.~Evans, and M.~Todisco, ``{An explainability study of the constant Q cepstral coefficient spoofing countermeasure for automatic speaker verification},'' in \emph{Speaker and Language Recognition Workshop (Odyssey)}, 2020.

\bibitem{salvi2023towards}
D.~Salvi, P.~Bestagini, and S.~Tubaro, ``Towards frequency band explainability in synthetic speech detection,'' in \emph{2023 31st European Signal Processing Conference (EUSIPCO)}.\hskip 1em plus 0.5em minus 0.4em\relax IEEE, 2023, pp. 620--624.

\bibitem{cozzolino2015splicebuster}
D.~Cozzolino, G.~Poggi, and L.~Verdoliva, ``{Splicebuster: A new blind image splicing detector},'' in \emph{IEEE International Workshop on Information Forensics and Security (WIFS)}.\hskip 1em plus 0.5em minus 0.4em\relax IEEE, 2015.

\bibitem{muller2021speech}
N.~M. M{\"u}ller, F.~Dieckmann, P.~Czempin, R.~Canals, K.~B{\"o}ttinger, and J.~Williams, ``Speech is silver, silence is golden: What do asvspoof-trained models really learn?'' in \emph{Conference of the International Speech Communication Association (INTERSPEECH)}, 2021.

\bibitem{mari2022sound}
D.~Mari, F.~Latora, and S.~Milani, ``The sound of silence: Efficiency of first digit features in synthetic audio detection,'' in \emph{2022 IEEE International Workshop on Information Forensics and Security (WIFS)}.\hskip 1em plus 0.5em minus 0.4em\relax IEEE, 2022, pp. 1--6.

\bibitem{defossez2020real}
A.~D{\'e}fossez, G.~Synnaeve, and Y.~Adi, ``Real time speech enhancement in the waveform domain,'' in \emph{Conference of the International Speech Communication Association (INTERSPEECH)}, 2020.

\bibitem{reddy2020interspeech}
C.~K. Reddy, V.~Gopal, R.~Cutler, E.~Beyrami, R.~Cheng, H.~Dubey, S.~Matusevych, R.~Aichner, A.~Aazami, S.~Braun \emph{et~al.}, ``The interspeech 2020 deep noise suppression challenge: Datasets, subjective testing framework, and challenge results,'' in \emph{Conference of the International Speech Communication Association (INTERSPEECH)}, 2020.

\bibitem{subakan2021attention}
C.~Subakan, M.~Ravanelli, S.~Cornell, M.~Bronzi, and J.~Zhong, ``Attention is all you need in speech separation,'' in \emph{IEEE International Conference on Acoustics, Speech and Signal Processing (ICASSP)}, 2021.

\bibitem{Maciejewski2020WHAMR}
M.~Maciejewski, G.~Wichern, and J.~Le~Roux, ``Whamr!: Noisy and reverberant single-channel speech separation,'' in \emph{IEEE International Conference on Acoustics, Speech and Signal Processing (ICASSP)}, May 2020.

\bibitem{todisco2019asvspoof}
M.~Todisco, X.~Wang, V.~Vestman, M.~Sahidullah, H.~Delgado, A.~Nautsch, J.~Yamagishi, N.~Evans, T.~Kinnunen, and K.~A. Lee, ``{ASVspoof 2019: Future horizons in spoofed and fake audio detection},'' in \emph{Conference of the International Speech Communication Association (INTERSPEECH)}, 2019.

\bibitem{yamagishi2021asvspoof}
J.~Yamagishi, X.~Wang, M.~Todisco, M.~Sahidullah, J.~Patino, A.~Nautsch, X.~Liu, K.~A. Lee, T.~Kinnunen, N.~Evans \emph{et~al.}, ``{ASVspoof 2021: accelerating progress in spoofed and deepfake speech detection},'' in \emph{Automatic Speaker Verification and Spoofing Countermeasures Challenge}, 2021.

\bibitem{ravanelli2018speaker}
M.~Ravanelli and Y.~Bengio, ``{Speaker Recognition from Raw Waveform with SincNet},'' in \emph{IEEE Spoken Language Technology Workshop (SLT)}, 2018.

\bibitem{muller2022does}
N.~M. M{\"u}ller, P.~Czempin, F.~Dieckmann, A.~Froghyar, and K.~B{\"o}ttinger, ``Does audio deepfake detection generalize?'' in \emph{Conference of the International Speech Communication Association (INTERSPEECH)}, 2022.

\bibitem{reimao2019dataset}
R.~Reimao and V.~Tzerpos, ``For: A dataset for synthetic speech detection,'' in \emph{International Conference on Speech Technology and Human-Computer Dialogue (SpeD)}.\hskip 1em plus 0.5em minus 0.4em\relax IEEE, 2019.

\bibitem{yi2022add}
J.~Yi, R.~Fu, J.~Tao, S.~Nie, H.~Ma, C.~Wang, T.~Wang, Z.~Tian, Y.~Bai, C.~Fan \emph{et~al.}, ``{ADD 2022: the First Audio Deep Synthesis Detection Challenge},'' in \emph{IEEE International Conference on Acoustics, Speech and Signal Processing (ICASSP)}, 2022.

\end{thebibliography}

\end{document}